# Characterization and manipulation of intervalley scattering induced by an individual monovacancy in graphene

Yu Zhang[1,2,3,#,*], Fei Gao[4,#], Shiwu Gao[5], Mads Brandbyge[4], and Lin He[3,*]

[1]School of Integrated Circuits and Electronics, MIIT Key Laboratory for Low-Dimensional Quantum Structure and Devices, Beijing Institute of Technology, Beijing 100081, China.

[2]Advanced Research Institute of Multidisciplinary Sciences, Beijing Institute of Technology, Beijing 100081, China.

[3]Center for Advanced Quantum Studies, Department of Physics, Beijing Normal University, 100875 Beijing, China.

[4]Center for Nanostructured Graphene, Department of Physics, Technical University of Denmark, DK-2800 Kongens Lyngby, Denmark.

[5]Beijing Computational Science Research Center, 100193, Beijing, China.

[#]Y.Z and F.G. contributed equally to this work.

[*]Correspondence and requests for materials should be addressed to Yu Zhang (e-mail: yzhang@bit.edu.cn) and Lin He (e-mail: helin@bnu.edu.cn).

**Intervalley scattering involves microscopic processes that electrons are scattered by atomic-scale defects on nanometer length scales. Although central to our understanding of electronic properties of materials, direct characterization and manipulation of range and strength of the intervalley scattering induced by an individual atomic defect have so far been elusive. Using scanning tunneling microscope, we visualized and controlled intervalley scattering from an individual monovacancy in graphene. By directly imaging the affected range of intervalley scattering of the monovacancy, we demonstrated that it is inversely proportional to the energy, i.e., it is proportional to the wavelength of massless Dirac Fermions. A giant electron-hole asymmetry of the intervalley scattering is observed because that the monovacancy is charged. By further charging the monovacancy, the bended electronic potential around the monovacancy softened the scattering potential, which, consequently, suppressed the intervalley scattering of the**

**monovacancy.**

Exploring the exact nature of emergent physical phenomena induced by atomic defects requires experimentally measurements at the nanoscale. One representative example is the local magnetism induced by monovacancy in graphene. Early transport and magnetic measurements of mesoscopic samples with amount of vacancies reported either Kondo effect or spin-half paramagnetism, confirming the local magnetism of the monovacancy [1,2]. However, the averaged magnetic moment per monovacancy was measured to be only about 0.1 $\mu_B$, which is much smaller than ~1.5 $\mu_B$ predicted in theory [3]. Later, nanoscale spectroscopy studies revealed different magnetic states and Kondo screening of the local magnetism of the monovacancy in graphene [4-6], helping to understand the much smaller magnetic moment of the monovacancy in mesoscopic measurements. Moreover, measurements on individual atomic defect may reveal results beyond that in mesoscopic measurements. For example, recent experiments demonstrated that the local wavefront dislocations in Friedel oscillations induced by an individual atomic impurity could reveal Berry phase signatures of the system [7-9].

Besides the local magnetism, the most relevant electronic property of the monovacancy in graphene is the intervalley scattering. In two-dimensional hexagonal materials, such as graphene, the valley index adds to the carrier's charge and spin, enabling additional degree of freedom and providing the opportunity of information coding and processing [10-14]. However, the atomic defects are supposed to generate intervalley scattering [15-32], which greatly limits the ability to manipulate the valley degree of freedom. Previous studies on the intervalley scattering induced by atomic defects mainly relied on transport measurements of mesoscopic graphene samples [25-36]. However, the scattering centers in transport measurements are always random and disordered, and their types and charge states are indistinguishable, owing to the lack of spatial resolution. In such a case, many seemingly contradictory phenomena were observed in transport measurements, including weak-localization/weak-antilocalization [25,26] and electron-hole symmetry/asymmetry effects [32,35,36], which need further exploration. In this Letter, we report nanoscale probing of the intervalley scattering induced by an individual monovacancy in graphene via scanning tunneling microscopy (STM) and spectroscopy (STS) measurements. A monovacancy in graphene locally introduces low-energy $V_\pi$ states. The charge states of monovacancy can be reversibly tuned by changing the electron filling of the $V_\pi$ state and precisely monitored by the Landau level spectroscopy. Our experiment indicates that the affected range of intervalley scattering by the monovacancy is proportional to the wavelength of massless Dirac fermions in graphene. A giant electron-hole asymmetry of the intervalley scattering is observed because that the monovacancy is charged. By further charging the monovacancy, the scattering potential around the monovacancy is softened and, consequently, the intervalley scattering of the monovacancy is suppressed. Our results provide an in-depth understanding of rich transport phenomena from a microscopic point of view.

In our experiments, a large-area multilayer graphene with a high density of monovacancies was directly synthesized on Ni foils by using a chemical vapor deposition (CVD) method [6], and then, was transferred onto a $SiO_2$/Si wafer (see methods and Fig. S1 of the Supplemental Material [37]). There is usually a rotational misalignment between the graphene layers, giving rise to moiré patterns in topological STM images. Figure 1(a) shows a representative STM image of a monovacancy in the topmost graphene layer of twisted bilayer graphene (TBG) with the twist angle of about 6°. In such a case, the two graphene layers are electronically decouple from each other, leaving the topmost graphene layer behaviors as a nearly freestanding single layer graphene (SLG) [38-44]. This result can be further verified by the π Berry phase signature and Landau level spectroscopy of massless Dirac Fermions in our STM and STS measurements, as we will demonstrate later.

The atomic-resolution STM image of a monovacancy in the AB-stacked region of TBG exhibits a distinctive triangular $\sqrt{3} \times \sqrt{3}$ $R$ 30° interference pattern (Fig. 1(a)), which is similar to that in SLG [4-6,45,46]. In fact, the location of the monovacancy has almost no influence on the studied phenomena in our current work, owing to the electronically decoupling of the topmost graphene sheet [38-44]. Figure 1(b) shows the fast Fourier transform (FFT) image of Fig. 1(a). The outer/inner six bright spots are the reciprocal lattice of the topmost SLG/TBG, and the middle six bright spots located at K and K′ valleys arise from the monovacancy-induced intervalley scattering. According to the inverse FFT by filtering a pair of K and K′ valleys, we observe a robust $N = 2$ additional wavefronts around the monovacancy, as shown in Fig. 1(c). Such a phenomenon is attributed to the rotation of sublattice pseudospin during the monovacancy-induced intervalley scattering process, and is a signature of π Berry phase in this system [7-9]. Therefore, our experiments demonstrate that the topmost graphene layer in large-angle TBG behaviors as a SLG.

Now we explore the electronic properties around an individual monovacancy in graphene under perpendicular magnetic fields $B$. Figures 1(d) and 1(e) show spatially resolved STS spectra recorded across the monovacancy under $B = 0$ T and 8 T, respectively (the spatial-resolved STS spectra recorded under other magnetic fields are given in Fig. S2 of the Supplemental Material [37]). Away from the monovacancy, the STS spectra exhibit a typical "V" shape under $B = 0$ T, together with the Landau quantization of massless Dirac fermions under $B = 8$ T, implying the two graphene layers in large-angle TBG are electronically decoupled [39-44]. As approaching to the monovacancy, the low-energy $V_\pi$ state gradually appears and is expected to split into two spin-polarized states, which is similar to that in the SLG [3,5]. In the following, we mainly focus on the energy of the $V_\pi$ state. We find that the $V_\pi$ state generated by a monovacancy in TBG usually lies at several tens of meV below the Dirac point $E_D$, as marked in Figs. 1(d) and 1(e). Such a result is quite different from that in the SLG where the $V_\pi$ state is just at $E_D$ [3].

To understand the phenomenon that monovacancy-induced $V_\pi$ state in the TBG is slightly away from the $E_D$, we carry out density functional theory (DFT) calculations. Figure 2(a) shows the optimized atomic structure of monovacancy in TBG with the twist angle of about 6°, and the monovacancy is placed at the AB-stacked region. Our calculations indicate that the monovacancy-induced $V_\pi$ state in the TBG always lies below the $E_D$ (Fig. 2(b)), and is almost independent of the atomic sites of the monovacancy. Such results are well consistent with our experiments. To gain further insights into the energy-separated $V_\pi$ state and $E_D$, we calculate the spatially resolved charge density difference $\Delta_\rho$ around the monovacancy, which can intuitively reflect the monovacancy-induced interlayer charge transfer in the TBG (more details are given in Supplemental Material [37]). From Fig. 2(c), we can observe an obvious electron transfer from the underlying layer into the topmost layer of TBG in the vicinity of the monovacancy, which is expected to be the root cause of the energy-separated $V_\pi$ state and $E_D$.

Now we concentrate on the spatial distribution of the monovacancy-induced intervalley scattering in TBG under $B = 0$ T. Figures 3(a) and (b) show two typical STM images recorded under $B = 0$ T at the sample bias of -0.2 V and -0.5 V, respectively. The range of the monovacancy-induced intervalley scattering $d$ can be defined from the real-space STM images as the distance from the site of missing carbon atom to the approximate scattering termination point where the $\sqrt{3} \times \sqrt{3}$ interference patterns are absent, as marked in Figs. 3(c) and (d), respectively (more details are given in Figs. S3 and S4 [37]). As the energy approaching to $E_D$, the length of monovacancy-induced intervalley scattering significantly increases and extends to the whole detected regions of $15 \times 15$ nm$^2$ STM images when reaching $E_D$, as shown in Fig. S5 [37]. Here we summarize the electron energy versus $1/d$ measured in our experiment in Fig. 3(e) and a linear relationship can be clearly observed. Taking into account the linear dispersion of the charge carriers in graphene, our result indicates that the affected range of intervalley scattering by an individual monovacancy is proportional to the wavelength of massless Dirac fermions in graphene. Therefore, the intervalley scattering of mesoscopic graphene samples with a certain density of monovacancies can be much enhanced when the Fermi energy is near $E_D$. This result explains the origin of the transition from weak localization to weak antilocalization in graphene when the energy of carriers is tuned away from $E_D$ in previous transport measurements [47]. According to our experiment, an obvious electron-hole asymmetry of the intervalley scattering, i.e., the slope for *E* vs 1/*d* is quite different for the electron and hole sides, is clearly observed. Such a giant electron-hole asymmetry is attributed to the fact that the monovacancy is charged, thus resulting in the surrounding massless Dirac fermions scattered more strongly when they are attracted to the charged defect than when they are repelled from it.

Theoretically, the charge state of the monovacancy in graphene can be tuned by the doping of the monovacancy-induced $V_\pi$ state. In our experiments, the charge state of the monovacancy can be reversibly tuned by perpendicular magnetic fields. Figure 4(a) shows representative STS spectra acquired at the monovacancy as a function of magnetic field *B* from 0 T to 8 T with the variation of 0.5 T. Apart from the evolution of the Landau levels as massless Dirac Fermions, the monovacancy-induced $V_\pi$ state becomes more localized as increasing the magnetic field. Moreover, the $V_\pi$ state trends to monotonously shift from -7 meV under *B* = 0 T towards a lower energy, and reaches the minimum of -26 meV under *B* = 8 T, as marked by $V_\pi$ states in Fig. 4(a). Such *B*-dependent $V_\pi$ states, on the one hand, is attributed to the redistribution of electronic states of graphene beneath the STM tip that are generated by the confinements of tip-induced electronic potentials and external magnetic fields (Fig. S6 [37]), as demonstrated recently [48-57], on the other hand, is attributed to the *B*-dependent electron transfers between the adjacent two graphene sheets around the defect, based on our STS measurements (Fig. 1) and DFT calculations (Fig. 2).

Theoretically, a monovacancy in graphene is able to host a local positive charge [41,58], and such charge state is not necessarily integer valued due to a wide energy width of the monovacancy-induced $V_\pi$ state [59]. As shown in Fig. 4(b,c), the monovacancy is expected to be neutral when the $V_\pi$ state is fully filled, and is gradually charged with the $V_\pi$ state shifting across the Fermi energy [34]. In our experiments, the charge states of a monovacancy under magnetic fields can be precisely monitored by the spatial-resolved Landau level spectroscopy. As shown in Fig. 1(e), we can observe an energy downshift of *N* = 0 LL, $\Delta E_{0\ LL}$, in the vicinity of the monovacancy relative to its value far away. The value of $\Delta E_{0\ LL}$ is derived from the screening of a charged monovacancy and can be expressed as $\Delta E_{0\,LL} \approx -(Z/\kappa)[e^2/(4\sqrt{2\pi}\varepsilon_0 l_B)]$. Here $\varepsilon_0$ is the permittivity of free space, $e$ is the electron charge, $l_B = \sqrt{\hbar/(eB)}$ is the magnetic length, $\hbar$ is the reduced Planck constant, and $\kappa$ is the effective dielectric constant [41,59,60]. Therefore, the charge state $Z$ of the monovacancy in graphene under various magnetic fields can be precisely extracted, as summarized in Fig. 4(d). Obviously, the charge state of the monovacancy shows an approximately monotonous decreasing from $Z \approx 0.5$ to 0.2 as the magnetic field increases from $B = 1.5$ T to 8.0 T, and can be extrapolated to $Z \approx 0.6$ when $B = 0$ T.

The intensity of monovacancy-induced intervalley scattering in graphene can be strongly influenced by the charge states. By charging the monovacancy, the scattering potential around the monovacancy can be softened from the atomically sharp scattering potential to the long-range scattering potential, and consequently, the intervalley scattering of the monovacancy can be efficiently suppressed (Fig. 4(c)). Figure 4(e) shows typical line profiles along the directions of G-Γ-G′ and K-Γ-K′ in the FFT images, which is acquired from the STS maps of a monovacancy in graphene under $B = 8$ T when $V_{\text{bias}} = 0.5$ V (Fig. S6 [37]). Here we summarize the intensity ratio of intervalley scattering $I_K$ and reciprocal lattice $I_G$, $I_K/I_G$, as a function of the charge state $Z$ in Fig. 4(f). As $Z$ decreasing, $I_K/I_G$ gradually enhances, together with the increasing spatial extension range of $\sqrt{3} \times \sqrt{3}$ $R$ 30° interference patterns in real space. Therefore, our results explicitly verify that the relative intensity of monovacancy-induced intervalley scattering can be strongly suppressed when the monovacancy is charged, as schematically shown in Fig. 4(c). In addition, we should emphasize that the rotation of pseudospin during the intervalley scattering process remains unchanged under different charge states.

In summary, we direct image the intervalley scattering induced by an individual monovacancy in graphene. We find that the affected range of intervalley scattering by the monovacancy is proportional to the wavelength of massless Dirac fermions in graphene, accompanied by a giant electron-hole asymmetry. By further charging the monovacancy, the scattering potential around the monovacancy is softened and, consequently, the intervalley scattering of the monovacancy is suppressed. Our results pave the way to control the intervalley scattering in multivalley systems, promoting the development of valleytronics.

**Acknowledgements**

This work was supported by the National Key R and D Program of China (Grant Nos. 2021YFA1401900, 2021YFA1400100), National Natural Science Foundation of China (Grant Nos. 12141401,11974050) and the China Postdoctoral Science Foundation (Grant No. 2021M700407).

**References**

1. J. Chen, L. Li, W. G. Cullen, E. D.Williams, M. S. Fuhrer, Tunable Kondo effect in graphene with defects. *Nat. Phys.* **7,** 535-538 (2011).

2. R. R. Nair, M. Sepioni, I. Tsai, O. Lehtinen, J. Keinonen, A. V. Krasheninnikov, T. Thomson, A. K. Geim, I. V. Grigorieva, Spin-half paramagnetism in graphene induced by point defects. *Nat. Phys.* **8,** 199-202 (2012).

3. O. V. Yazyev, L. Helm, Defect-induced magnetism in graphene. *Phys. Rev. B* **75,** 125408 (2007).

4. Y. Jiang, P. Lo, D. May, G. Li, G. Guo, F. B. Anders, T. Taniguchi, K. Watanabe, J. Mao, E. Y. Andrei, Inducing Kondo screening of vacancy magnetic moments in graphene with gating and local curvature. *Nat. Commun.* **9,** 2349 (2018).

5. Y. Zhang, S. Li, H. Huang, W. Li, J. Qiao, W. Wang, L. Yin, K. Bai, W. Duan, L. He, Scanning Tunneling Microscopy of the π Magnetism of a Single Carbon Vacancy in Graphene. *Phys. Rev. Lett.* **117,** 166801 (2016).

6. Y. Zhang, F. Gao, S. Gao, L. He, Tunable magnetism of a single-carbon vacancy in graphene. *Sci. Bull.* **65,** 194 (2020).

7. C. Dutreix, H. González-Herrero, I. Brihuega, M. I. Katsnelson, C. Chapelier, V. T. Renard, Measuring the Berry phase of graphene from wavefront dislocations in Friedel oscillations. *Nature* **574,** 219-222 (2019).

8. Y. Zhang, Y. Su, L. He, Intervalley quantum interference and measurement of Berry phase in bilayer graphene. *Phys. Rev. Lett* **125,** 116804 (2020).

9. Y. Zhang, Y. Su, L. He, Quantum Interferences of Pseudospin-Mediated Atomic-Scale Vortices in Monolayer Graphene. *Nano Lett.* **21,** 2526-2531 (2021).

10. A. Rycerz, J. Tworzydlo, C. W. J. Beenakker, Valley filter and valley valve in graphene. *Nat. Phys.* **3,** 172-175 (2007).

11. D. Xiao, W. Yao, Q. Niu, Valley-Contrasting Physics in Graphene: Magnetic Moment and Topological Transport. *Phys. Rev. Lett.* **99,** 236809 (2007).

12. D. Pesin, A. H. MacDonald, Spintronics and pseudospintronics in graphene and topological insulators. *Nat. Mater.* **11,** 409-416 (2012).

# Figures

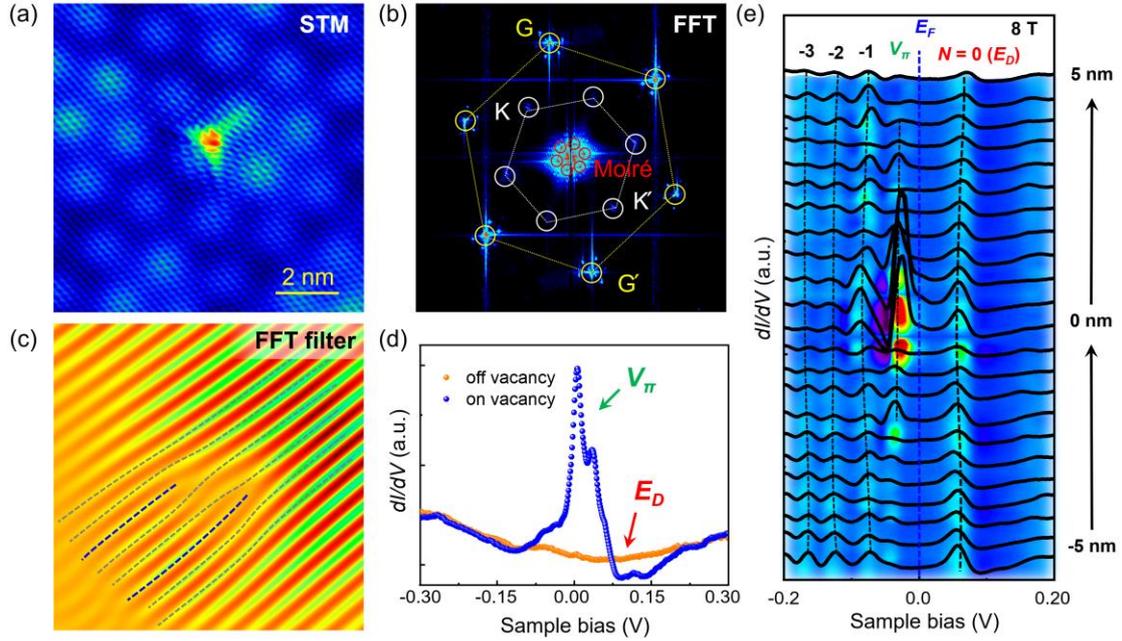

FIG. 1. (a) Atomic resolution STM image of a monovacancy in the topmost layer of TBG. (b) Corresponding FFT image of panel a. The outer and inner hexangular spots, marked by the yellow and red circles, correspond to the reciprocal lattice of the topmost graphene layer and the moiré pattern of TBG. The hexangular spots marked by the white circles correspond to the monovacancy-induced intervalley scattering process. (c) FFT-filtered STM image along a specific direction of the intervalley scattering by the time reversal. The $N = 2$ additional wavefronts are highlighted by the blue dashed lines. (d) Typical STS spectra recorded on and off the monovacancy in the absence of magnetic fields. The Dirac point $E_D$ and the monovacancy-induced spin-polarized states $V_\pi$ are marked in the panel. (e) The evolution of STS spectra recorded across the monovacancy under the perpendicular magnetic field of 8 T. The monovacancy locates at 0 nm. The Landau level indices are marked in the panel.

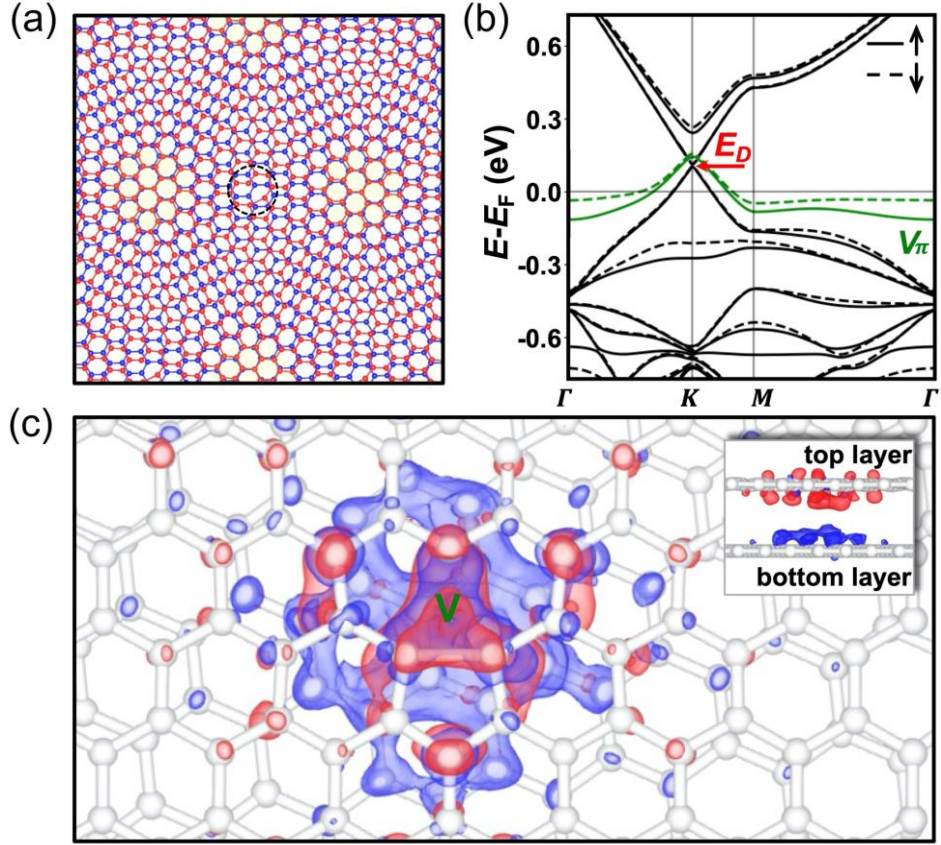

FIG. 2. (a) Atomic structure of TBG with the twist angle of 6°. A monovacancy is located at the topmost layer of TBG in the AB-stacked region, as marked in the center of the black circle. The red and blue spheres indicate C atoms of the topmost and underlying graphene layers, respectively. (b) Band structure of a monovacancy in the AB-stacked region of TBG. The $V_\pi$ states induced by the monovacancy are marked by the green lines. The energy of Dirac point is marked by the red arrow. Spin-up and spin-down electronic states are marked by the solid and dashed lines, respectively. (c) Charge density difference $\Delta_\rho$ around the monovacancy in TBG with the isosurface value of 0.0015 e/Å$^3$. Red and blue clouds indicate the electron accumulation and depletion, respectively.

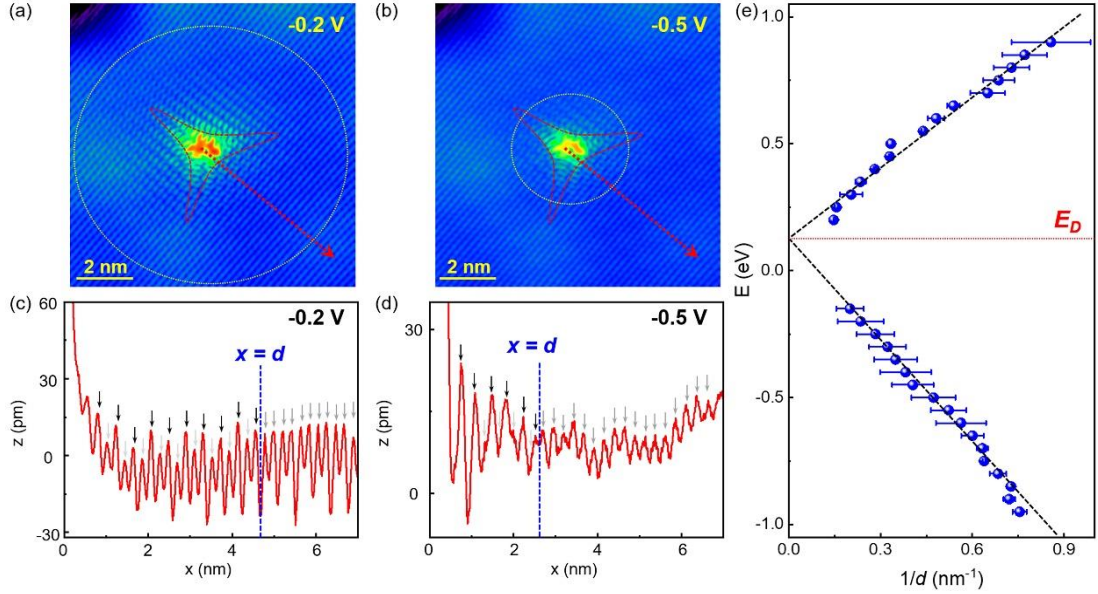

FIG. 3. (a,b) Atomic resolution STM image of a monovacancy in the topmost layer of TBG acquired at the sample bias of -0.2 V and -0.5 V with $512 \times 512$ pixels, respectively. The yellow dashed circles, demonstrating the ranges of defect-induced intervalley scattering, are guides to the eye. (c,d) Line profiles of the red arrows in panels a and b. The arrows indicate the locations of the carbon atoms in the topmost graphene layer. The alternate heights marked by black and light gray arrows for $x < d$ imply the existence of signal from the intervalley scattering, while the homogeneous heights marked by dark gray arrows for $x > d$ imply the negligible signal from the intervalley scattering. The $x = d$ is the precise length of the intervalley scattering, determined as the position when the difference of the apparent heights between two adjacent carbon atoms is less than 1 pm or the apparent heights no longer follow the rule of alternation. (e) A linear relation between electron energy ($E - E_D$) and $1/d$.

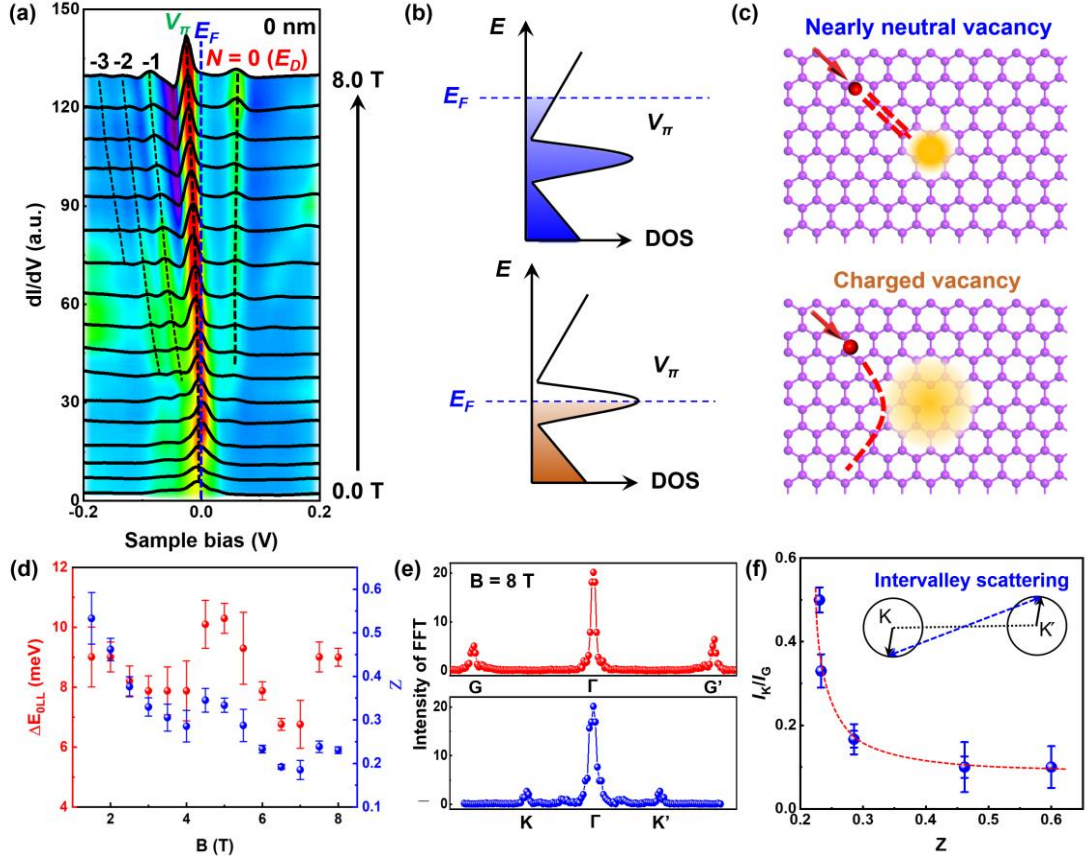

FIG. 4. (a) The evolution of STS spectra recorded at the monovacancy as a function of perpendicular magnetic field $B$ from 0 to 8 T as a variation of 0.5 T. The $V_\pi$ states and the Landau level indices of massless Dirac fermions are labeled in the panel. The Fermi energy is marked by the blue dashed line. (b) Schematic DOS at the monovacancy in TBG. Under $B = 0$ T, the $V_\pi$ state is partially filled, implying the monovacancy is charged. As increasing the perpendicular magnetic field, the $V_\pi$ state is gradually filled and finally is fully filled under $B = 8$ T, implying the monovacancy is nearly neutral. (c) Schematic scattering of carriers induced by the nearly neutral and charged monovacancy. (d) The energy shift of $N = 0$ LL, $\Delta E_{0LL}$, and the charge state $Z$ of the monovacancy as a function of magnetic fields. (e) Line profiles along the directions of G-Γ-G′ and K-Γ-K′ in the FFT images, which is acquired from the STS maps of a monovacancy in TBG under $B = 8$ T when $V_{bias} = 0.5$ V. (f) The intensity ratio of intervalley scattering and reciprocal lattice in FFT images, $I_K/I_G$, as a function of the charge state $Z$. Inset: schematic of the intervalley scattering.